\documentclass[draft,grl]{agu2001}
\usepackage{lineno}
\usepackage{graphicx}
\authorrunninghead{GEORGOULIS}
\titlerunninghead{MAGNETIC COMPLEXITY AND ERUPTIVE ACTIVITY IN THE SUN}
\authoraddr{Manolis K. Georgoulis, 
The Johns Hopkins University Applied Physics Laboratory, 11100 Johns
Hopkins Rd., Laurel, MD 20723, USA. (manolis.georgoulis@jhuapl.edu)}
\begin{document}

%
%
%
\setkeys{Gin}{draft=false}
\title{Magnetic Complexity in Eruptive Solar Active Regions and 
Associated Eruption Parameters}
\author{Manolis K. Georgoulis}
\affil{The Johns Hopkins University Applied Physics Laboratory, 
Laurel, Maryland, USA}
\begin{abstract}
Using an efficient magnetic complexity index in the active-region
solar photosphere, we quantify the 
preflare strength of the photospheric magnetic
polarity inversion lines in 
23 eruptive active regions with flare/CME/ICME
events tracked all the way from the Sun to the Earth. We find
that active regions with more intense polarity inversion lines
host statistically stronger flares and faster, more impulsively
  accelerated, CMEs. No
significant correlation is found between the strength of the
inversion lines and the flare soft X-ray rise times, 
the ICME transit times, and the peak $D_{st}$ indices
of the induced geomagnetic storms. Corroborating these and previous
results, we speculate on a possible interpretation for the connection 
between source active regions, flares, and CMEs. Further work is needed to 
validate this concept and uncover its physical details. 
\end{abstract}
\begin{article}
\section{Introduction}
The Solar and Heliospheric Observatory (SoHO) has established coronal
mass ejections (CMEs) as an integral part of solar eruptions. 
Today we know that there exist slow and fast CMEs associated with eruptive
quiet-Sun filaments and solar flares, respectively, 
with a wide velocity distribution
ranging between a few tens of $km/s$ to $\sim 3000\;km/s$
(St. Cyr et al. 2000; Yashiro et
al. 2004; Yurchyshyn et al. 2005). Solar flares are an exclusive
characteristic of solar active regions (ARs), so fast CMEs, typically
with velocities $\ge 750\;km/s$, are active-region CMEs (Sheeley et
al. 1999). Furthermore, the soft X-ray rise phase of the eruptive
flares almost coincides with the main
acceleration phase of the resulting CMEs (Zhang et al. 2001; Zhang and
Dere 2006). 

Observations and models have been utilized to  
relate specific AR characteristics with CME speeds. 
Following a ``big AR syndrome'',
extraordinary ARs can trigger superfast CMEs (Gopalswamy et al. 2005), 
but a more quantitative connection is lacking. 
Only recently, Qiu and Yurchyshyn (2005) 
reported a strong correlation between CME speeds and the  
reconnected magnetic flux in two-ribbon flares, Su et al. (2007)
combined magnetic flux and shear to improve correlations with flare
magnitudes and CME speeds, and T\"{o}r\"{o}k and Kliem (2007) 
concluded that increased magnetic complexity, reflected on steep
magnetic gradients in the source ARs' corona, tends 
to produce faster CMEs. 

The above studies suggest that complex (multipolar and/or with
pronounced magnetic polarity inversion lines [PILs]) ARs 
tend to produce faster CMEs. Here, we investigate this effect further,
reporting on work done in the framework of the
Living With a Star Coordinated Data Analysis Workshops (LWS/CDAW). 
We identified 23 source ARs with eruptions 
{\it unambiguously} traced from the solar surface to 1 AU. For these
ARs, we correlate the {\it peak} photospheric complexity prior to an
eruption with various eruption parameters, 
including flare magnitudes, plane-of-sky 
CME velocities, and (assumed constant) CME acceleration magnitudes.
Our analysis involves full-halo CMEs with source ARs located close to
disk center. Therefore, it should be kept in mind that 
significant discrepancies
may exist between the measured (plane-of-sky) and the true
(unprojected) CME velocities (Schwenn et al. 2005), that may have an impact on
correlations. For this and other reasons, as discussed below,  
we emphasize the statistical aspect of the reported correlations,
trying to avoid strong quantitative conclusions. 
\section{Magnetic Complexity Analysis and Data Selection}
Georgoulis and Rust (2007) defined the {\it effective connected
magnetic field strength}, $B_{eff}$, that characterizes 
a given AR at a given time. The larger the value of $B_{eff}$, the
more intense the PIL(s) present in the AR. The intensity of a PIL
increases when massive amounts of bipolar magnetic flux are
tightly concentrated around it. To calculate
$B_{eff}$ for the $p$ positive-polarity and $n$ negative-polarity flux
concentrations that comprise an AR, we calculate two $p \times n$
connectivity matrices: $\Phi _{i,j}$, containing the magnetic fluxes
that connect a positive-polarity 
concentration $i$ ($i \in [1,p]$) to a negative-polarity 
concentration $j$ ($j \in [1,n]$), and $L _{i,j}$, containing the
respective separation lengths of the connections. 
Then, $B_{eff}$ is the sum of all finite 
elements $(\Phi _{i,j} / L_{i,j}^2)$. 

\begin{figure}
 \noindent\includegraphics[width=39pc]{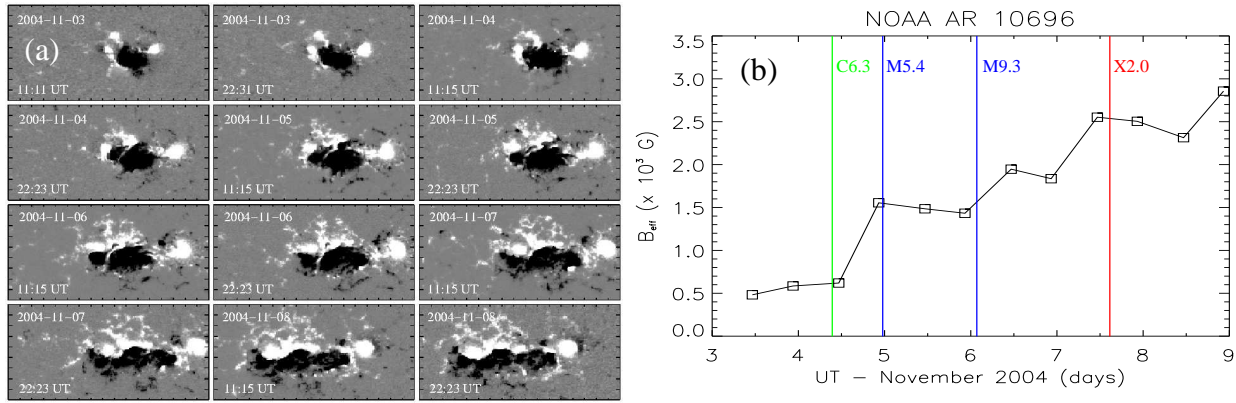}
 \caption{Photospheric evolution in NOAA AR 10696  over a period of 6
 days in 2004 November. (a) Normal photospheric magnetic field in the
 AR. Notice the gradual enhancement of the PIL's strength and spatial extent. 
 Tic mark separation in all images is $10$'' ($\sim 7250\;km$ in 
 the solar photosphere). Solar north is up; west is to the right.
 (b) Timeseries of the respective  $B_{eff}$-values. Notice the
 repeated flaring activity (onset times indicated by the color lines) as
 $B_{eff}$ increases.}
 \label{f1}
 \end{figure}

Georgoulis and Rust (2007) calculated $B_{eff}$ for $\sim 2140$ SoHO/MDI
magnetograms corresponding to 298 ARs. To avoid severe projection
effects, each AR was required to be located 
within $41^o$ EW from solar disk
center. Each AR needed 6-7 days to traverse this $82^o$-zone, and this
determined the typical observing period. 
To further account for projection effects, we 
(i) estimated the normal AR magnetic field by dividing the
line-of-sight field by $cos \theta$, where $\theta$ is the angular
distance of each location from disk center, 
(ii) constructed the local, heliographic, plane, and 
(iii) interpolated the normal field 
on the heliographic plane. We found that $B_{eff}$
efficiently distinguishes eruptive from non-eruptive ARs
{\footnote{In passing, we note that other magnetic complexity
    measures introduced for this purpose include, e.g., the fractal
    dimension of ARs (McAteer et al. 2005), the length of the main
    PIL (Falconer et al. 2006 and references therein), and the magnetic
    flux along the PIL (Schrijver 2007).}} 
and that a well-defined probability for major (X- and M-class, although
{\it not} their exact magnitudes) flares depends solely on $B_{eff}$. 
An example of how an increase in 
$B_{eff}$ translates to stronger PILs that, in turn, give rise to 
repeated flaring activity, is shown in Figure \ref{f1}.  

Zhang et al. (2007), on the other hand, identified the 
solar sources of 88 geoeffective ($D_{st} \le -100\;nT$)
solar eruptions. Each of these eruptions was traced from solar 
source to geomagnetic effect. An AR source could be identified 
for 54 of the above 88 events, with 23 different source ARs present. 

Combining the above two works, we calculated the peak
  pre-eruption (12 hr, at most, before each flare's onset)
  $B_{eff}$-values for these 23 source ARs. The AR and eruption data
  are summarized in Table \ref{tb1}.
\section{Results}
We seek a possible link between the peak AR photospheric complexity, 
quantified by the peak $B_{eff}$, and the flare
magnitude or CME kinematics. 
The flare magnitudes are related to the peak  
preflare $B_{eff}$-values in Figure \ref{f3}a. To calculate the
logarithmic flare magnitude we arbitrarily assign a magnitude of 1 to
a C1.0 flare. 
This implies a magnitude of 10, 100, and 1000, for a M1.0, X1.0, and
X10, flares, respectively (dashed lines). 
Notice the significant correlation coefficients (cc) 
between the flare magnitude and $B_{eff}$, despite the
scatter. The correlation clearly implies an increasing flare
magnitude for an increasing preflare $B_{eff}$. 
However, predictions of the flare magnitude using the shown scaling
are not recommended. This is because an AR may not yield its strongest
flare within the typical 6-7 day observing period despite showing a
high (close to peak) $B_{eff}$-value.
The scatter in Figure \ref{f3}a might also highlight 
the probabilistic (or even stochastic) nature of flare triggering,
with the flare magnitude depending in part 
on the local magnetic conditions and 
their synergy. The goodness of fit in Figure \ref{f3}a has 
a confidence level of $\sim 98.6$\%. {\it In brief, 
increasing $B_{eff}$ in an AR statistically implies 
stronger flares triggered in the AR}.  

In Figure \ref{f3}b we correlate the plane-of-sky 
CME velocities $V_{CME}$ with the
peak preflare $B_{eff}$-values. Although the small dynamical range of
$V_{CME}$ gives rise to lower and more fragile 
correlation coefficients, 
the scatter around the least-squares best fit appears smaller in
this case. This allows the introduction of a scaling relation between
$V_{CME}$ and peak preflare $B_{eff}$ that reads 
\begin{equation}
V_{CME} (km/s) \simeq 87.3 \times B_{eff}^{0.38 \pm 0.12} (G)\;\;.
\label{eq1}
\end{equation}
Equation (\ref{eq1}) is, again, not recommended for accurate
predictions of $V_{CME}$ but it suggests that {\it more complex ARs},
with larger and more intense PILs, statistically give rise to 
{\it faster} CMEs. For the smallest considered $B_{eff}$-value 
($\sim 500\;G$) we anticipate $V_{CME} \simeq 920\;km/s$, 
in good agreement with the slower end of AR CMEs 
($\sim 750\;km/s$). For our largest $B_{eff}$ ($\sim 5000\;G$), 
we expect $V_{CME} \simeq 2200\;km/s$, that is relatively close to 
(but somewhat smaller than) the largest measured CME velocities 
($\sim 2800\;km/s$). The difference between the 
two extreme velocities is 
$\sim 21$\%, with a typical $\sim 10$\%-uncertainty for the measured
plane-of-sky CME speed (J. Zhang, private communication). 
The goodness of fit in Figure \ref{f3}b has 
a confidence level of $\sim 91$\%. 
{\it Therefore, it is clear that increasing $B_{eff}$ in an AR 
statistically implies faster CMEs triggered in the AR.} 

In Figure \ref{f3}c we correlate the CME acceleration magnitude
$\gamma _{CME}$ with the peak preflare $B_{eff}$. 
We have estimated a constant 
$\gamma _{CME}$ by the ratio $(V_{CME}/T_{flare})$, 
i.e., by following the ``flare-proxy'' approach (Zhang and Dere 2006)
implying that $T_{flare}$ also reflects the main CME acceleration
phase. Although the trend in Figure \ref{f3}c is to attain stronger 
$\gamma _{CME}$ with increasing $B_{eff}$, the correlation is not as
appreciable as in Figures \ref{f3}a, \ref{f3}b. The goodness of the fit 
is also lower (confidence level $\sim 80$\%). Besides numerical effects
(e.g., discrepancies between plane-of-sky and true CME velocity), the
weaker correlation between $\gamma _{CME}$ and $B_{eff}$ may be due
to (i) the assumption of a constant CME acceleration magnitude, 
and/or (ii) the very weak anti-correlation between $T_{flare}$ and $B_{eff}$  
(cc $\in (-0.14, -0.18)$ - not shown). The
loose association between $T_{flare}$ and $B_{eff}$ means that 
the intensity of PILs in ARs correlates only weakly 
with the impulsiveness of the flares triggered in these ARs.
{\it Nevertheless, Figure \ref{f3}c suggests 
that increasing $B_{eff}$ in an AR statistically implies more
impulsively accelerated CMEs triggered in the AR.} 

We did not find significant correlations (maximum cc $\in (0.2, 0.5)$ and 
maximum confidence levels $\sim 70$\%) 
when correlating $B_{eff}$ with (i) the ICME transit time
$T_{ICME}$ and (ii) the peak absolute $D_{st}$ index of the resulting
geomagnetic storms. Neither result is surprising: 
Schwenn et al. (2005) and Manoharan (2006) report some correlation
between $T_{ICME}$ and $V_{CME}$ but with substantial scatter. Other
heliospheric effects may also impact the velocity profile , and hence the
arrival time, of ICMEs (Chen 1996; Cargill 2004; Tappin 2006).
Regarding $D_{st}$, many factors, besides the
source AR's complexity, affect the ICME geoeffectiveness. 
These factors include, but are not
limited to, the CME's source location in the solar disk, 
the orientation of the possible post-eruption flux rope, 
in-situ heliospheric distortions, turbulence, interactions
with other heliospheric transients, and the ICME velocity profile.
\section{Conclusions and Discussion}
In a previous work (Georgoulis and Rust 2007) we 
quantified the photospheric magnetic complexity in solar ARs, where
complexity reflects the strength of PILs in these
ARs. In this work we combined our sample of ARs with the
sample of AR sources that triggered major geomagnetic eruptions 
(Zhang et al. 2007). We identified 23 source ARs for which 
we have preflare AR magnetograms, and we correlated the
peak pre-eruption AR complexity with various eruption parameters. 

Significant correlations were uncovered when plotting the peak value of the AR
complexity index, $B_{eff}$, vs. (from stronger to weaker correlation) 
(i) the flare magnitude, (ii) the CME velocity, and (iii) an assumed constant
CME acceleration magnitude. Though not ideal for predictive purposes, these
correlations clearly imply that more complex ARs, with intense PILs,
can produce statistically stronger flares and 
faster, more impulsively accelerated, CMEs. Our finding
goes a step further than the usual ``big AR syndrome'': big 
(flux-massive) ARs are not necessarily ARs with intense PILs, 
and hence with large $B_{eff}$-values, although
the opposite is almost always true. 

\begin{figure}
 \noindent\includegraphics[width=20pc]{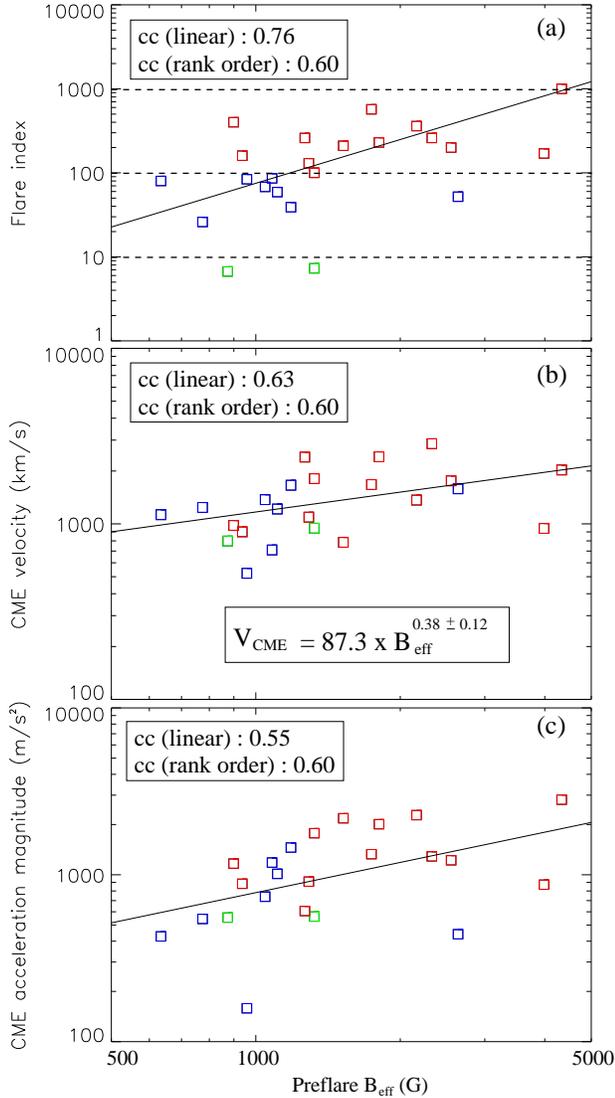}
 \caption{Correlation between the preflare $B_{eff}$-values and 
 (a) the  flare magnitude 
 (the three dashed lines indicate, from low to high,
 M1.0, X1.0, and X10. flares), (b) the plane-of-sky CME velocity, and 
 (c) the CME acceleration magnitude
 for the 23 events summarized in Table \ref{tb1}. In all plots, the
 straight lines indicate the least-squares best fit. C-, M-, and
 X-class flares are indicated by green, blue, and red squares,
 respectively. Both the Pearson (linear) 
 and the Spearman (rank order) correlation coefficients (cc) are shown.  
 For the correlation between $V_{CME}$ and $B_{eff}$ (b) the actual scaling
 formula is included in an inset.}
 \label{f3}
 \end{figure}

At this point, we can only speculate on a possible 
interpretation of our results, 
making it clear that only further work can validate or 
rule out our scenario. Given that the main
CME acceleration phase nearly coincides with the flare impulsive
phase, the initial ascending structure evolving into a CME 
(hereafter CME precursor) 
should start expanding {\it before} the flare. As it expands, the
CME precursor interacts with the surrounding PIL-supported magnetic
structure causing magnetic
reconnection. Reconnection in stronger magnetic fields organized along
more intense PILs statistically leads to stronger flares (Figure
\ref{f3}a). Stronger flares imply larger amounts of
released magnetic energy that, in turn, probably destabilize larger
parts of the PIL-sustained structure and/or accelerate the unstable
magnetic fields to higher speeds, thus giving rise to statistically
faster (Figure \ref{f3}b) and more impulsive (Figure \ref{f3}c) CMEs.  
Again, the above indicate only statistical trends - the
local field conditions at the flare location along the PIL as
well as the overlaying solar magnetic fields (steep magnetic
gradients [T\"{o}r\"{o}k and Kliem 2007], 
coronal null points [Antiochos et al. 1999], etc.) 
affect both flare magnitudes and CME velocity profiles.
Sometimes, fast CMEs are associated with relatively weak
flares or vice-versa (Vr\v{s}nak et al. 2005). 

The above idea might also help understand the difference
between (i) confined and eruptive flares and (ii) fast active-region
CMEs and slow quiet-Sun CMEs. In case the PIL-supported structure,
despite the reconnection, 
survives the perturbation applied by the CME precursor,  
a confined flare occurs{\footnote{Nindos and
Andrews (2004) suggest that the preflare magnetic helicity of the AR 
may determine whether a confined flare may occur.}}. If there is no
intense PIL, the CME precursor
might easily destabilize the surrounding magnetic structure 
(especially in the presence of ``open'' overlaying fields, such as 
a streamer in the high corona) but, since only minor reconnection is
expected, no major flare and a rather slow CME will occur. 
This might be the case for some quiet-Sun CMEs. 

While our statistical results appear solid, only 
further work can validate the above physical scenario. 
It would be essential to (i) uncover the physical mechanism(s)
responsible for the possible 
CME precursor (e.g., small-scale helical kink
instability, magnetic flux cancellation on the PIL, etc.), 
and (ii) find the relation between
the plane-of-sky velocity $V_{CME}$ and the true CME velocity and
determine whether this improves the correlation with $B_{eff}$ or the
source AR's complexity in general. The Solar Terrestrial Relations
Observatory (STEREO) can be instrumental in
revealing the true CME velocities to be used for 
definitive conclusions in our quest to understand the erupting Sun.
\begin{acknowledgments}
I am grateful to J. Zhang for many clarifying discussions and for his
help in handling and utilizing the data of Zhang et al. (2007). I also
thank A. Vourlidas for enlightening discussions in the physics of
CMEs, the organizers of the LWS/CDAW meetings for their sustained and
successful efforts, and the two anonymous referees, whose
thoughtful comments contributed significantly to this paper.  
This work has received partial support from NASA grant NNG05-GM47G. 
\end{acknowledgments}

\end{article}

%
\begin{table*}
\caption{Summary of the 23 LWS/CDAW events used in our study, where 
  the peak preflare $B_{eff}$-value of the source ARs 
  and the corresponding eruptions are selected. 
  Shown are the date and UT time of the flare onset, the flare class, 
  the soft X-ray flare rise time ($T_{flare}$), 
  the plane-of-sky CME velocity ($V_{CME}$), the
  CME acceleration magnitude ($\gamma _{CME})$, the ICME transit 
  time ($T_{ICME}$ - n/a means that $T_{ICME}$ could not be
  calculated), the peak $D_{st}$ index, the NOAA AR number, and
  the peak $B_{eff}$-value.}
\begin{tabular*}{\textwidth}{@{\extracolsep{\fill}}lcccccccccc}
\tableline
      &  \multicolumn{4}{c}{\it Flare} &
      \multicolumn{2}{c}{\it CME} & \multicolumn{2}{c}{\it ICME} &
      \multicolumn{2}{c}{\it Source} \\ 
      &  \multicolumn{4}{c}{\hrulefill} &
      \multicolumn{2}{c}{\hrulefill} & \multicolumn{2}{c}{\hrulefill}&
      \multicolumn{2}{c}{\hrulefill} \\ 
Event & Date & Onset & Class & $T_{flare}$ & $V_{CME}$ &
      $\gamma _{CME}$\tablenotemark{a} & $T_{ICME}$\tablenotemark{b} & $D_{st}$ & NOAA & $B_{eff}$ \\
      &   & (UT)  &       & ($min$)   & ($km/s$)  & 
      ($m/s^2$)      & ($hr$)     & ($nT$) & AR \# & ($G$)\\
\tableline
1 & 11/04/97  & 06:10 & X2.1 &6  & 785  & 2181 & 45.8 & -110 & 8100
& 1521.95\\
2 & 05/02/98  & 14:06 & X1.1 &11 & 938  & 1421 & n/a  & -205 & 8210
& 790.21\\
3 & 11/05/98  & 20:44 & M8.4 &55 & 523  & 158  & 79.3 & -142 & 8375
& 958.02\\
4 & 02/10/00  & 02:30 & C7.3 &28 & 944  & 562  & 54.5 & -133 & 8858
& 1323.7\\
5 & 07/14/00  & 10:54 & X5.7 &21 & 1674 & 1329 & 32.1 & -301 & 9077
& 1741.6\\
6 & 09/16/00  & 05:18 & M5.9 &20 & 1215 & 1013 & 39.7 & -201 & 9165
& 1108.7\\
7 & 10/09/00  & 23:50 & C6.7 &24 & 798  & 554  & 84.2 & -107 & 9182
& 873.81\\
8 & 11/26/00  & 17:06 & X4.0 &14 & 980  & 1167 & 46.9 & -119 & 9236
& 898.88\\
9 & 03/29/01 & 10:26 & X1.7 &18 & 942  & 872  & 42.6 & -387 & 9393
& 3985.6\\
10 & 04/10/01 & 05:30 & X2.3 &20 & 2411 & 2009 & 40.5 & -271 & 9415
& 1806.0\\
11 & 09/24/01 & 10:30 & X2.6 &66 & 2402 & 607  & n/a  & -102 & 9632
& 1265.2\\
12 & 10/19/01 & 16:50 & X1.6 &17 & 901  & 883  & 51.2 & -187 & 9661
& 937.13\\
13 & 10/25/01 & 15:26 & X1.3 &20 & 1092 & 910  & n/a  & -157 & 9672
& 1289.4\\
14 & 11/04/01 & 16:35 & X1.0 &17 & 1810 & 1775 & 44.4 & -292 & 9684
& 1324.1\\
15 & 04/17/02 & 08:26 & M2.6 &38 & 1240 & 544  & 63.6 & -149 &
9906 &  774.56\\
16 & 08/16/02 & 12:30 & M5.2 &60 & 1585 & 440  & 98.5 & -106 &
10069 & 2638.7\\
17 & 05/28/03 & 00:50 & X3.6 &10 & 1366 & 2277 & 36.2 & -144 &
10365 & 2162.4\\
18 & 10/29/03 & 20:54 & X10. &12 & 2029 & 2818 & 29.1 & -383 &
10486 & 4338.1\\
19 & 11/18/03 & 08:50 & M3.9 &19 & 1660 & 1456 & 49.2 & -422 &
10501 & 1183.5\\
20 & 07/20/04 & 13:31 & M8.6 &10 & 710  & 1183 & 52.5 & -101 &
10652 & 1080.9\\
21 & 11/07/04 & 16:54 & X2.0 &24 & 1759 & 1222 & 51.1 & -289 &
10696 & 2552.7\\
22 & 01/15/05 & 23:06 & X2.6 &37 & 2861 & 1289 & n/a  & -121 &
10720 & 2327.0\\
23 & 05/13/05 & 17:12 & M8.0 &44 & 1128 & 427  & 36.8 & -263 &
10759 & 634.30\\
\tableline
\end{tabular*}
\tablenotetext{a}{The CME acceleration magnitude is estimated 
  by the ratio $(V_{CME}/T_{flare})$.}
\tablenotetext{b}{The ICME transit time is calculated as the time
  difference between the ICME start time and the flare onset time} 
\label{tb1}
\end{table*}
%
%
%
\clearpage
%
%
\end{document}